\documentclass[a4paper,twocolumn,11pt,accepted=2024-06-01]{quantumarticle}
\pdfoutput=1
\usepackage[utf8]{inputenc}
\usepackage[english]{babel}
\usepackage[T1]{fontenc}
\usepackage{amsmath}
\usepackage{hyperref}
\usepackage[numbers]{natbib}
\usepackage{tikz}
\usepackage{amsfonts,color}
\usepackage{graphicx,dsfont}
\usepackage{physics}
\usepackage{soul}
\usepackage{xcolor}

\definecolor{orange}{RGB}{255, 165, 0}

\newlength{\ldag}
\begin{document}

\title{Quantum Phase Transitions in periodically quenched systems}

\author{Á. Sáiz}
\affiliation{Departamento de F\'isica At\'omica, Molecular y Nuclear, Facultad de F\'isica, Universidad de Sevilla, Apartado 1065, E-41080 Sevilla, Spain.}
\affiliation{Departamento de Física Aplicada III, Escuela Técnica Superior de Ingeniería, Universidad de Sevilla, E-41092 Sevilla, Spain}
\orcid{0000-0002-9378-3416}
\author{J. Khalouf-Rivera}
\affiliation{School of Physics, Trinity College Dublin, College Green, Dublin 2, Ireland}
\affiliation{Departamento de Física Aplicada III, Escuela Técnica Superior de Ingeniería, Universidad de Sevilla, E-41092 Sevilla, Spain}
\affiliation{Departamento de  Ciencias Integradas y Centro de Estudios Avanzados en F\'isica, Matem\'atica y Computaci\'on, Universidad de Huelva, 21071 Huelva, Spain}
\orcid{0000-0003-4201-2512}
\author{J. M. Arias}
\affiliation{Departamento de F\'isica At\'omica, Molecular y Nuclear, Facultad de F\'isica, Universidad de Sevilla, Apartado 1065, E-41080 Sevilla, Spain.}
\affiliation{Instituto Carlos I de F\'{\i}sica Te\'orica y Computacional, Universidad de Granada, Fuentenueva s/n, 18071 Granada, Spain }
\orcid{0000-0001-7363-4328}
\author{P. Pérez-Fernández}
\affiliation{Departamento de Física Aplicada III, Escuela Técnica Superior de Ingeniería, Universidad de Sevilla, E-41092 Sevilla, Spain}
\affiliation{Instituto Carlos I de F\'{\i}sica Te\'orica y Computacional, Universidad de Granada, Fuentenueva s/n, 18071 Granada, Spain }
\orcid{0000-0002-4706-7080}
\author{J. Casado-Pascual}
\affiliation{F\'isica Te\'orica, Universidad de Sevilla, Apartado de Correos 1065, Sevilla 41080, Spain} 
\email{jcasado@us.es}
\orcid{0000-0003-4118-5809}

\maketitle

\begin{abstract}
Quantum phase transitions encompass a variety of phenomena that occur in quantum systems exhibiting several possible symmetries. Traditionally, these transitions are explored by continuously varying a control parameter that connects two different symmetry configurations. Here we propose an alternative approach where the control parameter undergoes abrupt and time-periodic jumps between only two values. This approach yields results surprisingly similar to those obtained by the traditional one and may prove experimentally useful in situations where accessing the control parameter is challenging.
\end{abstract}

\section{Introduction}
Matter can exist in different structural states or phases that have different properties. The transitions between phases are characterized by rapid changes in the properties of the system. The study and characterization of phase transitions and critical phenomena is of great physical importance due, among other things, to the appearance of certain universal behaviors~\cite{PT_textbook,PT_textbook2}. Typically, the change from one phase to another is governed by a control parameter (temperature or other appropriate variable) and the phase transition is characterized by an order parameter that is zero in one phase and nonzero in the other. The way in which this change in the order parameter occurs makes it possible to classify the phase transitions~\cite{PT_textbook,PT_textbook2}.

More recently, the phenomenon of phase transitions has been extended to the quantum regime, giving rise to the so-called quantum phase transitions (QPTs)~\cite{Sachdev2011,Vojta_2003,Carr2010}.  For this, a physical system that can exhibit two different symmetries, ${\cal S}_1$ and ${\cal S}_2$, is usually considered. These symmetries are represented by two Hamiltonians, $H_1$ and $H_2$, respectively, that correspond to different structures and properties of the system. To study the transition from one of the symmetries to the other, a Hamiltonian of the form
\begin{equation}
	\label{Hamiltonian_CQPT}
 H = \xi H_2 + (1-\xi) H_1 
\end{equation}
is considered, where $\xi$ is a dimensionless parameter in the range $\xi \in [0,1]$ that serves as a control parameter. For $\xi=0$ and $\xi=1$ the system exhibits the symmetries ${\cal S}_1$ and ${\cal S}_2$, respectively, while for intermediate values of $\xi$ the symmetry is not well-defined and there is a competition between both symmetry phases. Various quantities, both static and dynamic, have been used to study and characterize the phase transition that occurs when the control parameter $\xi$ varies between $0$ and $1$. Specifically, it has been shown that certain properties of the ground state undergo abrupt changes for given intermediate critical values of $\xi$, giving rise to the so-called Ground-State Quantum Phase Transitions (GSQPTs)~\footnote{The acronym QPT has traditionally been used to denote ground-state quantum phase transitions. In the present work, QPT refers to quantum phase transitions in general.}. Excited-State Quantum Phase Transitions (ESQPTs) have also been reported, characterized by the appearance of  divergences in the level density at certain excitation energy values (see, e.g., Ref.~\cite{Cejnar2021} and references therein). Obviously, these divergences appear in the thermodynamic limit, where the number of particles tends to infinity. However, for small systems with finite number of particles, precursors of these divergences can be observed. 

In recent years, QPTs have been analyzed using time-periodic Hamiltonians and Floquet theory~\cite{FloquetPedroPRL, FloquetPedroPRA,rodriguez-vega2021,zhou_floquet_2021,deger_arresting_2022,zhao_probing_2022,PRB2022Jangjan}. However, most of the works have focused on the study of a new phenomenon, called Floquet Dynamical Quantum Phase Transitions (FDQPTs), characterized by the appearance of recurrent nonanalytical behaviors of certain quantities over time.  It is important to note that this phenomenon is different from the GSQPTs and ESQPTs mentioned above, which have been studied mainly using time-independent Hamiltonians. The question that naturally arises is whether GSQPTs and ESQPTs can also be analyzed using time-periodic Hamiltonians and Floquet theory. In this work, we show that the answer is yes, opening new avenues to study experimentally GSQPTs and ESQPTs  using  Floquet engineering.

The structure of the remainder of this work is as follows. In Section~\ref{Floquet-Theory}, the procedure developed to study QPTs using Floquet techniques is described. To facilitate readability, the mathematical validity conditions of this procedure are detailed in Appendix~\ref{app:condition}. In Section~\ref{PhysRea}, we investigate the application of this technique to a specific physical model, namely, the Lipkin-Meshkov-Glick (LMG) model~\cite{LIPKIN1965188,Meshkov1965199,Glick1965211}. The obtained results are analyzed in Section~\ref{Results}. For completeness, in Appendix~\ref{app:othermodels}, the proposed Floquet technique is also applied to a simple two-level model describing the interaction between individual atoms and diatomic homo-nuclear molecules. Finally, Section~\ref{Conclusions} presents a summary of the conclusions of our study

\begin{figure}
    \centering
    \includegraphics[scale=0.56]{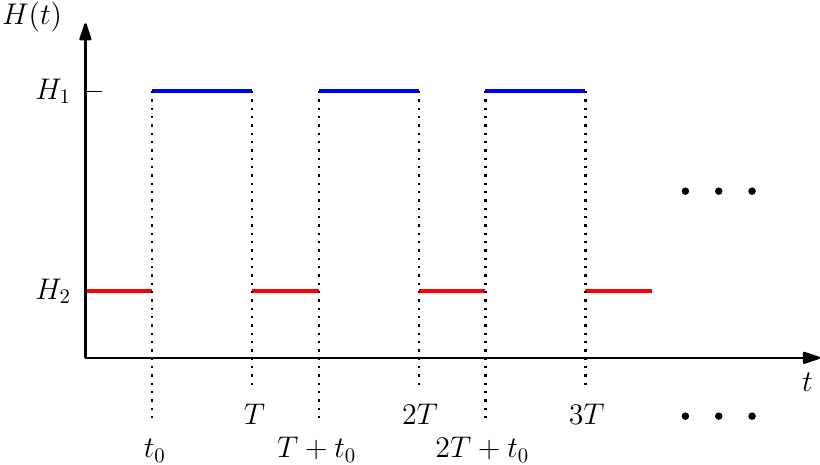}
    \caption{Sketch of the time-periodic Hamiltonian in Eq.~(\ref{T_D_Hamiltonian}). The quench time $t_0$ can be varied and serves as a control parameter. }
    \label{fig-1}
\end{figure}

\section{Description of Floquet techniques applied to QPTs}
\label{Floquet-Theory}
Let us consider a system characterized by a time-periodic Hamiltonian with period $T$ defined on $0\leq t<T$ by
\begin{equation}
	\label{T_D_Hamiltonian}
	H(t)=
	\begin{cases}
		H_2  &  \text{if } t \in [0,t_0) \\
		H_1 & \text{if } t \in [t_0,T),
	\end{cases}
\end{equation}
and extended periodically for all $t$ (see Fig.~\ref{fig-1}). As in the traditional QPTs, $H_1$ and $H_2$ are time-independent Hamiltonians with symmetries $\mathcal{S}_1$ and $\mathcal{S}_2$, respectively. The parameter $t_0$ (quench time) denotes the amount of time per period that the Hamitonian remains with the symmetry $\mathcal{S}_2$, and will be used as a control parameter. Obviously, for $t_0=0$ and $t_0=T$ the Hamiltonian is time-independent and equals to $H_1$ and $H_2$, respectively.

 Note that in the present case, the Hamiltonian has a  well-defined symmetry at all time instant, and the periodic transitions between symmetries $\mathcal{S}_1$ and $\mathcal{S}_2$ are discontinuous. By contrast, in the traditional QPTs the transition from $\mathcal{S}_2$ to $\mathcal{S}_1$ is continuous, since it is carried out by continuously varying the control parameter $\xi$ from $0$ to $1$. Furthermore, the Hamiltonian in Eq.~(\ref{Hamiltonian_CQPT}) only has a well-defined symmetry when the parameter $\xi$ takes the endpoints $0$ or $1$. Despite these important differences, surprisingly, the time-periodic Hamiltonian in Eq.~(\ref{T_D_Hamiltonian})  shows a very similar behavior to that observed in the traditional  QPTs [Eq.~(\ref{Hamiltonian_CQPT})], as is discussed below.

Due to the periodicity of the Hamiltonian in Eq.~(\ref{T_D_Hamiltonian}), Floquet theory~\cite{Grifoni1998a} can be used to analyze the time evolution of the system. According to Floquet's theorem, the Schr\"odinger equation corresponding to the Hamiltonian~(\ref{T_D_Hamiltonian}) possesses a complete set of solutions of the form $\smash{\ket{\Psi_j(t)}=e^{-i\epsilon_j t/\hbar}}\ket{\Phi_j(t)}$, with $j\in\{0,\cdots,D-1\}$, where $D$ is the dimension of the state space. The kets $\smash{\ket{\Phi_j(t)}}$ are $T$-periodic functions of time and are known as Floquet modes. The quantities  $\smash{\epsilon_j}$ are called quasienergies and can (and will) be taken to lie within the first Brillouin zone $\smash{[-\pi\hbar/T,\pi\hbar/T)}$.  Henceforth, we will assume that the quasienergies are labeled so that $\smash{\epsilon_{j+1}\geq\epsilon_j}$. 

The calculation of the Floquet modes $\smash{\ket{\Phi_j(t)}}$ and the quasienergies $\epsilon_j$ requires only the knowledge of the time-evolution operator $U(t,0)$  from $t=0$ to $t\in[0,T]$. Since the Hamiltonian in Eq.~(\ref{T_D_Hamiltonian}) is piecewise time-independent, $U(t,0)$ can be explicitly evaluated, yielding 
\begin{equation}
	\label{xidef}
	U(t,0)=
	\begin{cases}
		e^{-i t H_2/\hbar}  &  \text{if } t \in [0,t_0) \\
		 e^{-i (t-t_0) H_1/\hbar} 	e^{-i t_0 H_2/\hbar}& \text{if } t \in [t_0,T).
	\end{cases}
\end{equation}
The first step to calculate  $\smash{\ket{\Phi_j(t)}}$  and $\epsilon_j$ is to obtain the eigenvectors and eigenvalues of the operator
\begin{equation}
\smash{U(T,0)= e^{-i (T-t_0) H_1/\hbar} 	e^{-i t_0 H_2/\hbar}}
\label{UT0}.
\end{equation}
Since this operator is unitary, it possesses an orthonormal basis of eigenvectors. These eigenvectors provide the Floquet modes at $t=0$, i.e., the kets $\smash{\ket{\Phi_j(0)}}$. If $\lambda_j$ is the eigenvalue of $U(T,0)$ associated with the eigenvector $\smash{\ket{\Phi_j(0)}}$, the corresponding quasienergy  is given by
\begin{equation}
\epsilon_j=i \hbar \log(\lambda_j)/T,
\label{quasienergies}
\end{equation}
where $\log$ denotes the principal value of the logarithm.  Finally, the Floquet mode at any time $t$ can be obtained extending periodically the function 
\begin{equation}
	\label{FMt}
\ket{\Phi_j(t)}=e^{i t \epsilon_j/\hbar}U(t,0)\ket{\Phi_j(0)},
\end{equation}
with $U(t,0)$ given by Eq.~(\ref{xidef}).

It is worth mentioning that the quasienergy $\epsilon_j$ can be split into a dynamical and a geometrical contribution as $\smash{\epsilon_j=\bar{E}_j}-\hbar \varphi_j/T$, where
\begin{equation}
	\label{ME}
	\bar{E}_j=\frac{1}{T}\int_0^T dt \expval{H(t)}{\Phi_j(t)}
\end{equation}
is the mean energy, and
\begin{equation}
	\label{GP}
	\varphi_j=i \int_0^T dt \expval{\frac{d}{dt}}{\Phi_j(t)}
\end{equation}
is the geometric phase of the Floquet mode. From Eqs.~(\ref{xidef}), (\ref{FMt}), and (\ref{ME}), it is easy to see that the mean energy is given by
\begin{equation}
	\label{ME2}
	\begin{split}
\bar{E}_j = & \frac{t_0}{T}\expval{H_2}{\Phi_j(0)}\\ 
&+\frac{T-t_0}{T}\expval{e^{i t_0 H_2/\hbar}H_1e^{-i t_0 H_2/\hbar}}{\Phi_j(0)}.
\end{split}
\end{equation}
As will be seen below, the quantities $\epsilon_j$, $\bar{E}_j$, and $\varphi_j$ can be used to characterize the QPTs and ESQPTs described in this work. For this to be possible, the period $T$ must be taken sufficiently small compared to the characteristic time $\smash{T_\mathrm{c} =4\hbar/\sqrt{(E_{1,\mathrm{M}}-E_{1,\mathrm{m}})(E_{2,\mathrm{M}}-E_{2,\mathrm{m}})}}$, 
where $E_{j, \mathrm{M}}$ and $E_{j, \mathrm{m}}$ are, respectively, the maximum and minimum eigenvalues of $H_j$,  with $j=1,2$ (see Appendix~\ref{app:condition}). To avoid having to extend the spectrum of quasienergies to other Brillouin zones different from the first one, it is sufficient to additionally assume that $T/T_{\mathrm{c}}^{\prime}\leq \pi$, where $T_{\mathrm{c}}^{\prime} =\hbar/\max(\lvert E_{1,\mathrm{M}}\rvert, \lvert E_{1,\mathrm{m}}\rvert,\lvert E_{2,\mathrm{M}}\rvert, \lvert E_{2,\mathrm{m}}\rvert)$ is another characteristic time of the system (see Appendix~\ref{app:condition}).

\begin{figure}[ht]
    \centering
    \includegraphics[width=0.49\textwidth]{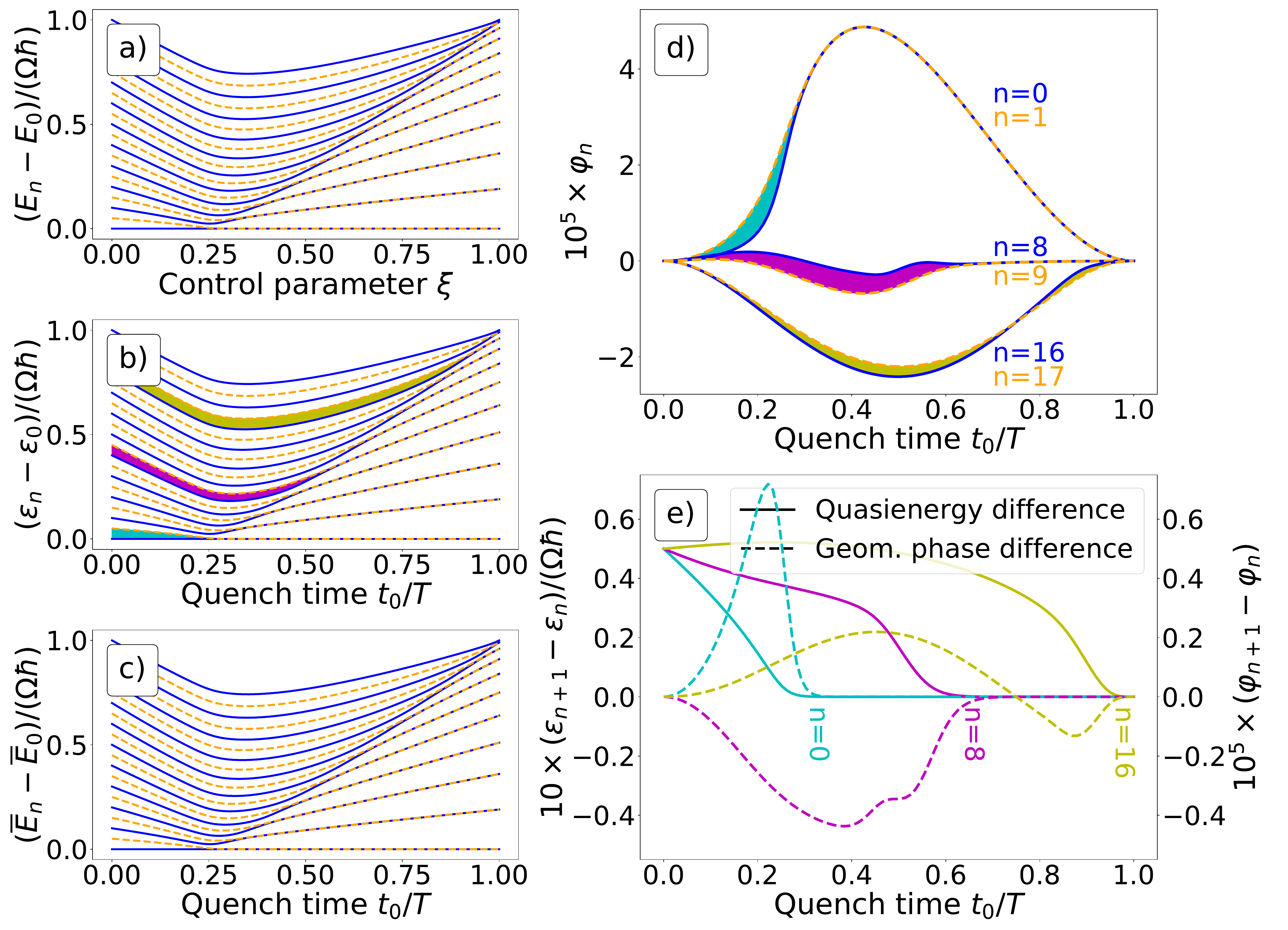}
    \caption{Panel~a) shows the dependence of the excitation energies $E_n-E_0$ on the control parameter $\xi$ for the time-independent Hamiltonian in Eq.~(\ref{Hamiltonian_CQPT}), with $H_1$ and $H_2$ given by the LMG model in Eq.~(\ref{eq:LMGH}).  The solid blue lines correspond to even $n$ and the dashed orange lines to odd $n$. Panels~b), c), d), and e) refer to the time-periodic Hamiltonian in Eq.~(\ref{T_D_Hamiltonian}), with $H_1$ and $H_2$ given by Eq.~(\ref{eq:LMGH}) and $T=1/\Omega$. Panels~b) and c) depict, respectively,  the differences of quasienergies, $\epsilon_n-\epsilon_0$, [calculated using Eq.~(\ref{quasienergies})] and of mean energies, $\smash{\bar{E}_n-\bar{E}_0}$,  [calculated using Eq.~(\ref{ME2})] as a function of the dimensionless quench time $t_0/T$, with the same line code as in panel~a). Panel~d) shows the geometric phases of the Floquet modes $n=0$, $1$, $8$, $9$, $16$, and $17$ as a function of $t_0/T$. Panel~e) depicts the differences $\epsilon_{n+1}-\epsilon_n$ (solid lines and left ordinate axis) and $\varphi_{n+1}-\varphi_n$ (dashed lines and right ordinate axis) as a function of $t_0/T$ for $n=0$, $8$, and $16$. These differences are highlighted in panels~b) and d). In all panels $N=20$.}
    \label{fig-2}
\end{figure}

\section{Physical realization}
\label{PhysRea}
In order to realize the ideas discussed in the preceding text, we will use a simple two-level model: the LMG model~\cite{LIPKIN1965188,Meshkov1965199,Glick1965211}. This model was originally proposed in the field of Nuclear Physics to validate many-body approximation methods and has subsequently been widely used in many different areas. 
In the field of phase transitions, the standard LMG Hamiltonian can be understood as a one-dimensional $N$-site spin-$1/2$ lattice with infinite interaction range~\cite{Richerme2014, Jurcevic2014}. The Hamiltonian can be written as Eq.~(\ref{Hamiltonian_CQPT}) with 
\begin{equation}
H_1= \frac{\Omega S_{z}}{2S}~~~{\rm and} ~~~ H_2 =-\frac{\Omega S_{x}^2}{\hbar S^2},
\label{eq:LMGH}
\end{equation}
where $\Omega$ is a constant with dimensions of frequency and collective spin operators $\smash{S_{\beta} = \sum_{i=1}^{N} s_{i,\beta}}$, for $\beta = x,y,z$, are introduced. Constant contributions to the Hamiltonian have been dropped. In this scheme, $S=N/2$ is the maximum collective spin. Concerning symmetries, the LMG Hamiltonian presents a $u(2)$ algebraic structure and possesses two dynamical symmetries: $u(2)\supset u(1)$ and $u(2)\supset so(2)$~\cite{Frankbook}. The symmetries $u(1)$ and $so(2)$ play, respectively, the role of ${\cal S}_1$ and ${\cal S}_2$ in our formalism, and are realized by the Hamiltonians ${H}_1$ and ${H}_2$  in Eq.~(\ref{eq:LMGH}), respectively.  Each of these symmetries is linked to a different structure (phase) of the system.

With the purpose of applying Floquet machinery to the LMG model, the Hamiltonian in Eq.~(\ref{T_D_Hamiltonian}) is considered, with $H_1$ and $H_2$ given by Eq.~(\ref{eq:LMGH}). Under this theoretical framework, the evolution operator $U(T,0)$ in Eq.~(\ref{UT0}), the Floquet modes in Eq.~(\ref{FMt}), the geometric phase of the Floquet modes in Eq.~(\ref{GP}), and the mean energy of the Floquet modes in Eq.~(\ref{ME2})  can be obtained by analytical expressions. The only numerical task is the diagonalization of the evolution operator $U(T,0)$.

\begin{figure}[ht]
\centering
\includegraphics[width=0.45\textwidth]{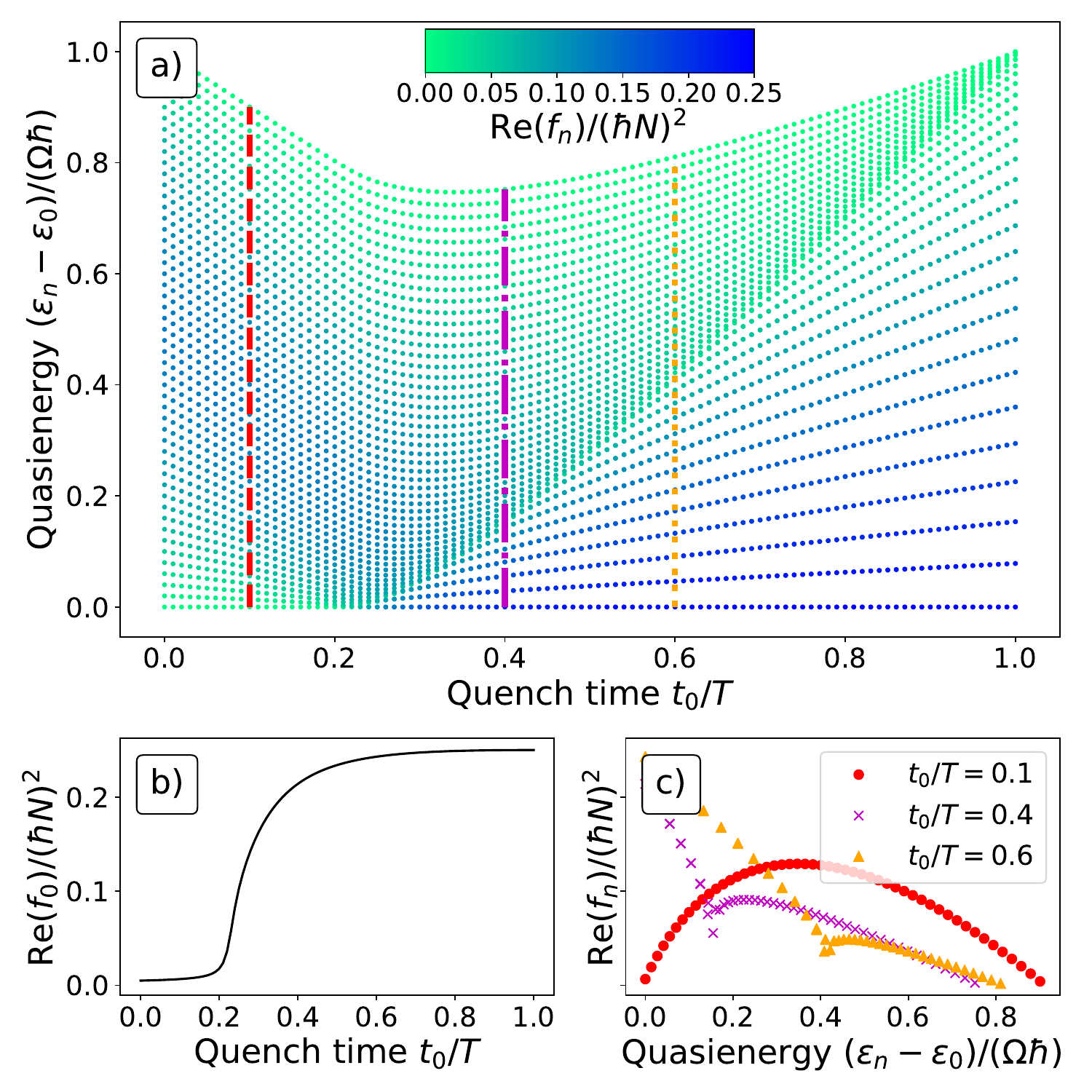}
\caption{Panel a) shows the same as Fig.~\ref{fig-2}.b) but for $N=50$ and using a color scale to indicate the value of the real part of the two-time correlator in Eq.~(\ref{TTC}). Panel b) shows the real part of this two-time correlator as a function of the dimensionless quench time $t_0/T$ for the Floquet mode with the lowest quasienergy in the first Brillouin zone. Panel c) shows the real part of this two-time correlator as a function of the quasi-energies $\epsilon_n-\epsilon_0$ for fixed quench times $t_0/T=0.1$ (red dots), $t_0/T=0.4$ (violet crosses), and $t_0/T=0.6$ (orange triangles). These values of the quench time are highlighted in panel a) with vertical red dashed, violet dot-dashed, and orange dotted  lines, respectively. In all panels, $T=1/\Omega$.}
\label{fig-3}
\end{figure}

\section{Results}
\label{Results}
The GSQPT and ESQPT appearing in the LMG model have been studied previously by varying the control parameter $\xi$ in Eq.~(\ref{Hamiltonian_CQPT}) (see Ref.~\cite{Cejnar2021} and references therein). This is depicted in Fig.~\ref{fig-2}.a),  where the excitation energies $E_n-E_0$ of the time-independent LMG Hamiltonian in Eqs.~(\ref{Hamiltonian_CQPT}) and (\ref{eq:LMGH})  are plotted with solid blue lines (even $n$) and dashed orange lines (odd $n$) as a function of the control parameter $\xi$, which  is varied continuously from $\xi=0$ [symmetry $u(1)$] to $\xi=1$ [symmetry $so(2)$]. Two phases are clear: in one [harmonic $u(1)$ symmetry limit] the excitation  energies are non-degenerated and equally spaced, while in the other [$so(2)$  symmetry limit] they are degenerated in pairs. For the GSQPT the transition is at $\xi=0.2$ in the large-$N$ limit~\cite{Dusuel2004,Romera2014}. The calculations presented here are for $S=N/2=10$, thus finite-$N$ effects are important. In addition, for the broken-symmetry phase a separatrix line of higher density of states is clearly observed, which marks the ESQPT. Different observables have been used to characterize both GSQPT and ESQPT always for the time-independent LMG model~\cite{Ribeiro2008}.

In order to determine if the GSQPT and ESQPT can also be identified using Floquet theory with the quench time $t_0$ as the control parameter, the time-periodic LMG Hamiltonian defined by Eqs.~(\ref{T_D_Hamiltonian}) and~(\ref{eq:LMGH}) is now considered. In this case, instead of the energies $E_n$, the Floquet quasienergies, $\epsilon_n$, and the mean energies of the Floquet modes, $\smash{\bar{E}_n}$, have been computed using Eqs.~(\ref{quasienergies}) and~(\ref{ME2}), respectively.  For this model, it can be easily seen that $T_{\mathrm{c}}^{\prime}=1/\Omega$ and,  if $S$ is even, $T_\mathrm{c} = 4/\Omega$ (see Appendix~\ref{app:condition}). 
For our calculations, we have chosen the value of the period to be $T=1/\Omega$,  so that $T/T_{\mathrm{c}}^{\prime}=1$ and $T/T_\mathrm{c} = 1/4$.
The results for the differences $\epsilon_n-\epsilon_0$ and $\smash{\bar{E}_n-\bar{E}_0}$ as a function of the dimensionless quench time $t_0/T$ are depicted in Figs.~\ref{fig-2}.b) and~\ref{fig-2}.c), respectively,  with the same line code as in Fig.~\ref{fig-2}.a). Remarkably,  Figs.~\ref{fig-2}.b) and~\ref{fig-2}.c) are almost indistinguishable from Fig.~\ref{fig-2}.a), showing that almost the same information about the GSQPT and ESQPT is obtained for the time-periodic system as for the time-independent Hamiltonian. This is noteworthy given that the period $T$ selected for the time-periodic Hamiltonian is only one quarter of the characteristic time $T_{\mathrm{c}}$. Although the values for the quasienergies and the mean energies shown in Figs.~\ref{fig-2}.b) and~\ref{fig-2}.c) are very similar to each other, they are not identical. To confirm this,  the dependence on $t_0/T$ of the geometric phases of the Floquet modes, $\smash{\varphi_n =(\bar{E}_n-\epsilon_n) T/\hbar}$,  is depicted in Fig.~\ref{fig-2}.d). For the sake of clarity only the modes $n=0$, $1$, $8$, $9$ $16$, and $17$ have been plotted. As can be seen, although the geometric phases of the Floquet modes are zero for $t_0=0$ and $t_0=T$, they are generally nonzero for intermediate values of $t_0$. Interestingly, the geometric phases of the Floquet modes become degenerate in pairs at approximately the same values of $t_0$ as the quasienergies, following the values along the separatrix line in Fig.~\ref{fig-2}.b). This is shown even more clearly in Fig.~\ref{fig-2}.e), where the differences $\epsilon_{n+1}-\epsilon_n$ (solid lines and left ordinate axis) and $\varphi_{n+1}-\varphi_n$ (dashed lines and right ordinate axis) are plotted versus the dimensionless quench time $t_0/T$ for $n=0$, $8$, and $16$.

From the above results, it seems clear that the information about the QPT contained in the time-independent Hamiltonian in Eq.~(\ref{Hamiltonian_CQPT}) is retained by the time-periodic Hamiltonian in Eq.~(\ref{T_D_Hamiltonian}) and reflected in the quasienergies, the mean energies, and the geometric phases of the Floquet modes. The question then arises whether it is possible to find an order parameter for the GSQPT and a good marker for the ESQPT.  To this end, here we propose the real part of the two-time correlator defined as
\begin{equation}
    f_j=\expval{S_x(T)S_x(0)}{\Phi_j(0)},
    \label{TTC}
\end{equation}
where $S_x(t)$  corresponds to the operator $S_x$ expressed in the Heisenberg picture. 

In Fig.~\ref{fig-3}, the time-periodic LMG Hamiltonian defined by Eqs.~(\ref{T_D_Hamiltonian}) and~(\ref{eq:LMGH}) is again considered but now for $N=50$. The dependence of the quasienergy differences $\epsilon_n-\epsilon_0$ on the dimensionless quench time $t_0/T$ is depicted in Fig.~\ref{fig-3}.a) with a color scale that indicates the value of the real part of the two-time correlator in Eq.~(\ref{TTC}).  In particular, for the Floquet mode corresponding to  the lowest quasienergy in the first Brillouin zone [horizontal line at $\epsilon_n-\epsilon_0=0$ in Fig.~\ref{fig-3}.a)], the dependence on $t_0/T$ of the real part of this correlator is depicted in  Fig.~\ref{fig-3}.b). As can be seen, the quantity $\mathrm{Re}(f_0)$ behaves as an order parameter, since it is zero in one phase and nonzero in the other, marking clearly the GSQPT around $t_0/T=0.2$. The possibility of detecting the ESQPT using the quantity $\mathrm{Re}(f_n)$ is explored in Fig.~\ref{fig-3}.c). For this purpose, three vertical lines are plotted in Fig.~\ref{fig-3}.a), which mark different quench times, specifically, $t_0/T=0.1$ (dashed red line), $0.4$ (dot-dashed violet line), and $0.6$ (dotted orange line). The dependence of $\mathrm{Re}(f_n)$ on the quasienergy difference $\epsilon_n-\epsilon_0$ along these three vertical lines is depicted in Fig.~\ref{fig-3}.c). For $t_0/T=0.1$ (red dots), $\mathrm{Re}(f_n)$ is a smooth function of $\epsilon_n-\epsilon_0$, which is consistent with the fact that for this quench time the system is at the same phase for any value $\epsilon_n-\epsilon_0$ [see Fig.~\ref{fig-3}.a)]. By contrast, for $t_0/T=0.4$ (violet crosses) and $t_0/T=0.6$ (orange triangles), abrupt changes are observed around $\epsilon_n-\epsilon_0=0.18\,\hbar\Omega$ and $\epsilon_n-\epsilon_0=0.42\,\hbar\Omega$, respectively, which are precisely the values at which the ESQPTs occur in Fig.~\ref{fig-3}.a). Therefore, we conclude that the real part of the two-time correlator in Eq.~(\ref{TTC}) can be used as a marker for both GSQPT and ESQPT, and provides relevant information on the phase diagram of the time-independent Hamiltonian in Eq.~(\ref{Hamiltonian_CQPT}) using the time-periodic Hamiltonian in Eq.~(\ref{T_D_Hamiltonian}).

The definition of the two-time correlator in Eq.~(\ref{TTC}) implicitly assumes that the system is prepared in a Floquet state at the initial time $t=0$. Given the potential challenges associated with implementing such initial conditions in practical applications, it would be advisable to expand the definition to encompass a broader range of initial conditions. Among other benefits, this expansion would enable us to evaluate the robustness of the resulting order parameter against variations in the initial condition.  With the purpose of extending the definition, we begin by initializing the system at time $t=0$ in an arbitrary state, which may be either pure or impure, characterized by the density operator $\rho(0)$.  Subsequently, the driving is initiated, and a stroboscopic measurement of the observable $[S_x(m T)S_x(0)+S_x(0)S_x(mT)]/2$  is conducted at time intervals $m T$, where $m$ represents any positive integer. The expected value of each of these measurements is determined by the real part of the expression $\mathrm{Tr}\left[S_x(m T) S_x(0) \rho(0)\right]$. Finally, the average of the outcomes over the number of periods $m$ for a sufficiently large maximum number of periods $M$ is computed, which is formally expressed as the real part of the limit
	\begin{equation}
		\label{AOP}
		\bar{f}=\lim_{M\to\infty}\frac{1}{M}\sum_{m=1}^{M} \mathrm{Tr}\left[S_x(m T)S_x(0)\rho(0)\right].
	\end{equation}

\begin{figure}[ht]
	\centering
	\includegraphics[width=0.45\textwidth]{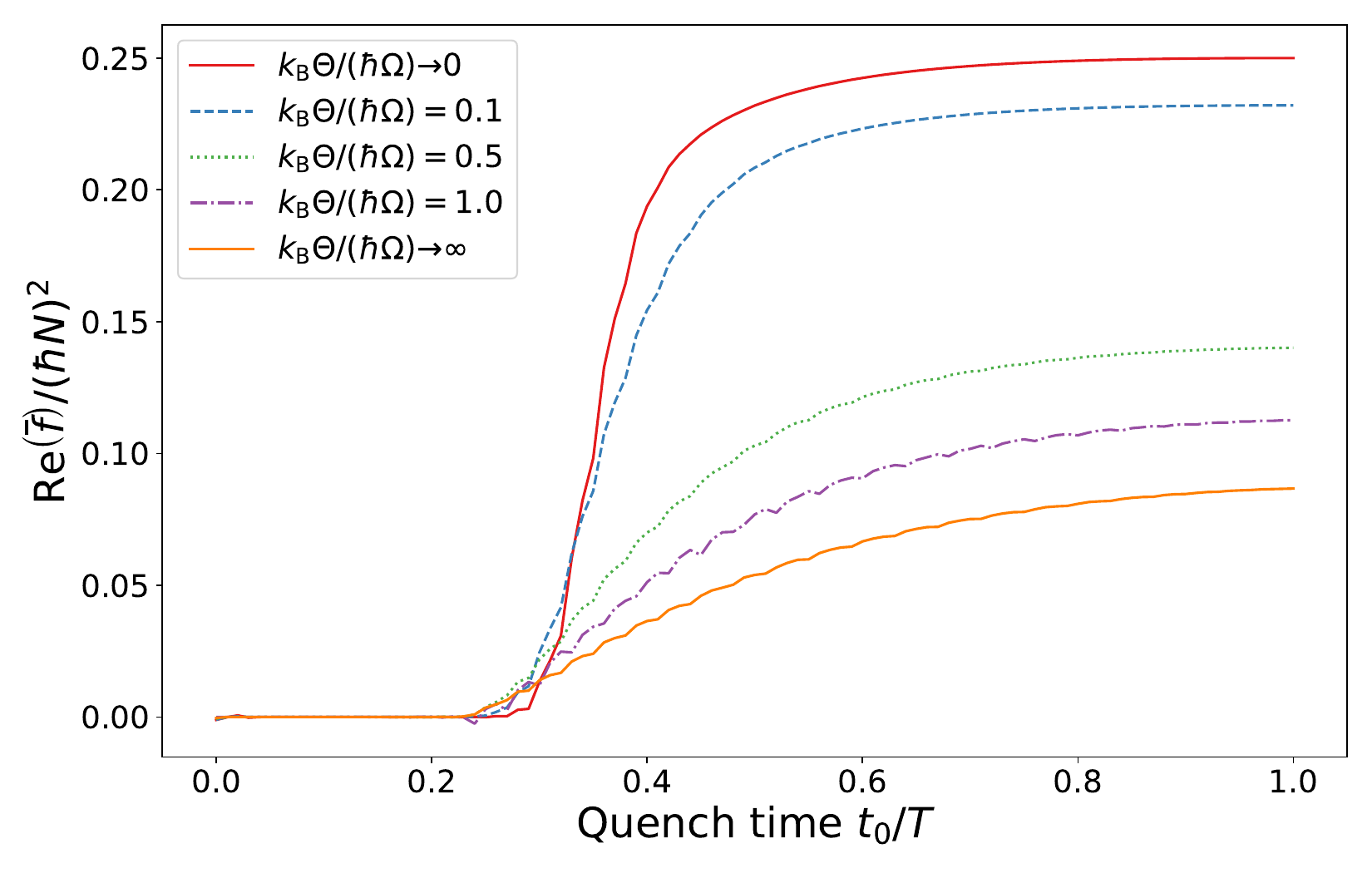}
\caption{ Dependence of the real part of the average two-time correlator, as defined by Eq.~(\ref{AOP}), on the dimensionless quench time $t_0/T$. The system is initially assumed to be at thermal equilibrium corresponding to the initial Hamiltonian $H_2$ at a specific absolute temperature $\Theta$, with varying values of the dimensionless temperature $k_{\mathrm{B}}\Theta/(\hbar \Omega)$. The parameters $N$ and $T$ remain consistent with those in Fig.~\ref{fig-3}, i.e., $N=50$ and $T=1/\Omega$. The limit in Eq.~(\ref{AOP}) was numerically evaluated using a maximum number of periods $M=10^4$.}
	\label{fig-4}
\end{figure}

To confirm the effectiveness of $\mathrm{Re}(\bar{f})$ as an order parameter and analyze potential thermal effects, we have followed the above procedure and evaluated $\mathrm{Re}(\bar{f})$ through an actual time-dependent simulation, initiated from various thermal equilibrium states corresponding to different temperatures.  Specifically, we have assumed that at the initial time $t=0$, the system has reached thermal equilibrium corresponding to the initial Hamiltonian $H_2$ at a certain absolute temperature $\Theta$. Thus, the initial density operator is $\rho(0)=e^{-H_2/(k_\mathrm{B}\Theta)}/Z(\Theta)$, where $k_{\mathrm{B}}$ denotes the Boltzmann constant, and $Z(\Theta)=\mathrm{Tr}[e^{-H_2/(k_\mathrm{B}\Theta)}]$ represents the partition function corresponding to the temperature $\Theta$. In Fig.~\ref{fig-4}, the results for $\mathrm{Re}(\bar{f})$ obtained using the aforementioned procedure for different temperatures are plotted against $t_0/T$. The values of $N$ and $T$ remain consistent with those in Fig.~\ref{fig-3}, i.e., $N=50$ and $T=1/\Omega$, and the numerical evaluation of the limit in Eq.~(\ref{AOP}) was conducted by selecting a maximum number of periods $M=10^4$. The limits $\Theta\to 0$ (red solid line) and $\Theta\to\infty$ (orange solid line) have been calculated using the initial states $\rho(0)=(\ketbra{g_2}{g_2}+\ketbra{g_2'}{g_2'})/2$ and $\rho(0)=I/(N+1)$, respectively, where $\{\ket{g_2},\ket{g_2'}\}$ represents an orthonormal basis of eigenvectors corresponding to the degenerate ground level of the Hamiltonian $H_2$, and $I$ denotes the identity operator. These initial states correspond to formally taking the limits $\Theta\to 0$ and $\Theta\to \infty$ of the canonical density operator $e^{-H_2/(k_\mathrm{B}\Theta)}/Z(\Theta)$.

As shown in Fig.~\ref{fig-4}, for all the considered temperature values, $\mathrm{Re}(\bar{f})$ exhibits a behavior similar to that observed in Fig.~\ref{fig-3} b), displaying a transition from a nearly zero value to a nonzero one at a certain value of $t_0/T$. The maximum value of $\mathrm{Re}(\bar{f})$ depends on the temperature considered, with a decrease observed as the temperature increases. The value of $t_0/T$ at which this transition occurs slightly depends on the temperature considered, but for all cases considered, it is above the value near $0.2$ observed in Fig.~\ref{fig-3} b). This increase may be attributed to the fact that while in Fig.~\ref{fig-3} b) the initial condition involved only the Floquet modes with the lowest quasienergy in the first Brillouin zone, all Floquet modes come into play for the curves in Fig.~\ref{fig-4}. Therefore, it is to be expected that the order parameter shown in Fig.~\ref{fig-3} b) more accurately reflects the exact value of $t_0/T$ at which the GSQPT takes place, while for the one depicted in Fig.~\ref{fig-4}, this value is slightly shifted to the right by effects due to the ESQPT. It is noteworthy that even at infinite temperature, $\mathrm{Re}(\bar{f})$ continues to behave as a reliable order parameter. Thus, the proposed averaged order parameter demonstrates robustness against potential heating induced by periodic driving. However, given the integrable nature of the system described by the Hamiltonian in Eq. (10), unlimited heating of the system is not anticipated~\cite{DAlessio2013}.

\section{Conclusions}
\label{Conclusions}
In this work, QPTs in a time-independent system are proposed to be studied by using a time-periodic Hamiltonian. This Hamiltonian exhibits two different symmetries and periodically switches from one symmetry to the other. Using the LMG model and through simple calculations in the framework of Floquet theory (mostly analytical except for a matrix diagonalization), we show that the relevant information about the phase diagram of the time-independent system is retained by the time-periodic Hamiltonian, provided the period $T$ is chosen to be sufficiently small compared to a characteristic time of the system. In particular, the GSQPT and ESQPT are clearly manifested on the quasienergies, mean energies, and geometric phases of the Floquet modes. Furthermore, the real part of the two-time correlator in Eq.~(\ref{TTC}) is shown to be a good marker not only for the GSQPT, but also for the ESQPT. Consequently, this type of time-dependent Hamiltonians with periodic quenches can be used to study and characterize the phase diagram and critical points of time-independent systems. This finding opens up new promising avenues for the use  of Floquet machinery in the field of QPT. In particular, our approach may be experimentally useful in situations where accessing the control parameter is challenging, as it only requires the use of two specific values of this parameter.

\begin{acknowledgments}
The authors gratefully acknowledge Rodrigo G. Corti\~{n}as for fruitful discussions and advice. In addition, we acknowledge project PID2022-136228NB-C22 funded by MICIU/AEI/10.13039/501100011033 and by ERDF, EU. We also acknowledge resources supporting this work provided by the CEAFMC and Universidad de Huelva High Performance Computer (HPC@UHU) funded by ERDF/MINECO project UNHU-15CE-2848. This work was partially supported by the Consejer\'{\i}a de Econom\'{\i}a, Conocimiento, Empresas y Universidad de la Junta de Andaluc\'{\i}a (Spain) under Groups FQM-160, FQM-177 and under projects P20-00617, P20-01247, and US-1380840 (FEDER); also through the projects PID2019-104002GB-C21, PID2019-104002GB-C22, and PID2020-114687GB-I00 funded by MCIN/AEI/10.13039/50110001103 and ``ERDF A way of making Europe''. JKR also acknowledges support from a Spanish Ministerio de Universidades "Margarita Salas" Fellowship. Furthermore, this work has been partially financially supported by the Ministry for Digital Transformation and of Civil Service of the Spanish Government through the QUANTUM ENIA project call-Quantum Spain project, and by the European Union through the Recovery, Transformation and Resilience Plan-NextGenerationEU within the framework of the Digital Spain 2026 Agenda.
\end{acknowledgments}


\appendix

\section{\label{app:condition} Applicability conditions of Floquet techniques for studying QPT}

As in Section~\ref{Floquet-Theory}, let us consider a system characterized by a time-periodic Hamiltonian with period $T$ defined on $0\leq t<T$ by Eq.~\eqref{T_D_Hamiltonian},
and extended periodically for all $t$. We will assume that the Hamiltonians $H_1$ and $H_2$ are bounded, and that $E_{j, \mathrm{M}}$ and $E_{j, \mathrm{m}}$ are, respectively, the maximum and minimum eigenvalues of $H_j$,  where $j=1,2$.

In order for the Floquet technique developed in Section~\ref{Floquet-Theory} to be applicable, it would suffice for the dynamics associated with the time-periodic Hamiltonian in Eq.~(1) to be similar to that of the time-independent Hamiltonian 
\begin{equation}
\label{S_Hamiltonian}
	H_{\mathrm{s}}=\left(1-\frac{t_0}{T}\right)H_1+\frac{t_0}{T} H_2
\end{equation}
for all values of $t_0$ in the interval $[0,T]$. 
In particular, the time-evolution operators from $t=0$ to $t=T$ should be similar for both Hamiltonians, i.e.,
 \begin{equation}
	e^{-i H_{\mathrm{s}} T /\hbar} \approx  e^{-i (T-t_0) H_1/\hbar} 	e^{-i t_0 H_2/\hbar}
\end{equation}
for all $t_0\in[0,T]$.

By considering the Baker-Campbell-Hausdorff formula~\cite{hall-2015}
\begin{equation}
	\begin{split}
		\log \left[ e^{-i (T-t_0) H_1/\hbar} 	e^{-i t_0 H_2/\hbar}\right]=& - \frac{i H_{\mathrm{s}} T }{\hbar}\\
		&+  \Delta(t_0)+\cdots,
		\label{BCH}
	\end{split}
\end{equation}
with 
\begin{equation}
	\label{defDelta}
	\Delta(t_0) =-\frac{(T-t_0)t_0}{2 \hbar^2}[H_1,H_2],
\end{equation}
it can be concluded that one  possible condition for both dynamics to be similar could be that $\Vert \Delta(t_0) \Vert$ is sufficiently small for all $t_0\in[0,T]$, where $\Vert \cdot \Vert$  denotes a certain matrix norm. Here, we will use the spectral norm $\Vert \cdot \Vert_2$ (see, e.g., Ref.~\cite{horn-1985}), which for a certain operator $A$ is defined as $\Vert A \Vert_2=\sqrt{\lambda_{\mathrm{M}}(A^{\dagger}A)}$, where $\lambda_{\mathrm{M}}(A^{\dagger}A)$ is the largest eigenvalue of $A^{\dagger}A$. 

It should be noted that since any operator commutes with another proportional to the identity operator $I$, the Hamiltonians $H_1$ and $H_2$ in Eq.~(\ref{defDelta}) can be replaced without changing the value of $\Delta(t_0)$ by $H_1(\alpha_1)=H_1+\alpha_1 I$ and $H_2(\alpha_2)=H_2+\alpha_2 I$, respectively, where $\alpha_1$ and $\alpha_2$ are two arbitrary constants. This change also does not affect the following terms of the Baker-Campbell-Hausdorff formula in Eq.~(\ref{BCH}), since these can also be expressed as commutator nestings of the operators $H_1$ and $H_2$~\cite{hall-2015}. As we will see later, the values of $\alpha_1$ and $\alpha_2$ will be chosen in such a way that the upper bound for $\Vert \Delta(t_0) \Vert_2$ is as low as possible. 

Taking into account that the spectral norm is sub-additive, i.e., $\Vert A+B \Vert _2\leq \Vert A \Vert_2+ \Vert B \Vert_2$, and sub-multiplicative, i.e., $\Vert AB \Vert_2 \leq \Vert A \Vert_2 \Vert B \Vert_2$, and given that $(T-t_0)t_0\leq T^2/4$ for all $t_0\in[0,T]$, it is easy to see that
\begin{equation}
\label{inequality1}
\lVert \Delta(t_0) \rVert_2 \leq \frac{T^2	\lVert H_1 (\alpha_1)\lVert_2\lVert H_2(\alpha_2) \lVert_2}{4 \hbar^2}.
\end{equation}
From the definition of the spectral norm, it follows that $\Vert H_j(\alpha_j)\Vert_2=\max(\abs{E_{j,\mathrm{M}}-\alpha_j},\abs{E_{j,\mathrm{m}}-\alpha_j})$, for $j=1,2$. Therefore, the smallest values of $\Vert H_j(\alpha_j)\Vert_2$ are attained when $\alpha_j=(E_{j,\mathrm{M}}+E_{j,\mathrm{m}})/2$, and are equal to $(E_{j,\mathrm{M}}-E_{j,\mathrm{m}})/2$. Substituting these values into Eq.~(\ref{inequality1}), we obtain that
\begin{equation}
	\label{inequality2}
	\lVert \Delta(t_0) \rVert_2 \leq \left(\frac{T}{T_\mathrm{c}}\right)^2,
\end{equation} 
where the characteristic time
\begin{equation}
\label{tc}
T_\mathrm{c}=\frac{4\hbar}{\sqrt{(E_{1,\mathrm{M}}-E_{1,\mathrm{m}})(E_{2,\mathrm{M}}-E_{2,\mathrm{m}})}}
\end{equation}
has been introduced. According to all the aforementioned, a condition for the Floquet techniques developed in Section~\ref{Floquet-Theory} to be applicable to the study of quantum phase transitions is that the period $T$ is sufficiently small compared to the characteristic time $T_{\mathrm{c}}$ in Eq.~(\ref{tc}).

\begin{figure*}
    \centering
    \includegraphics[width=0.9\linewidth]{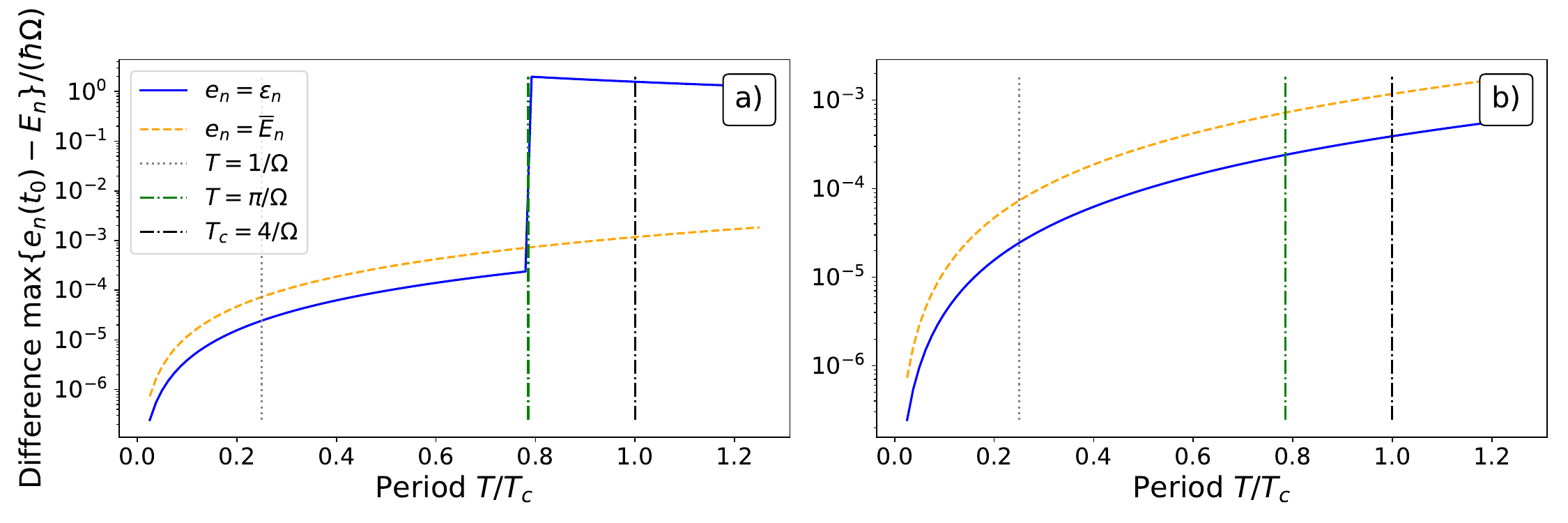}
    \caption{Dependence on the dimensionless period $T/T_{\mathrm{c}}$ of the maximum deviations between the results obtained for the time-periodic Hamiltonian $H(t)$ [Eq.~(\ref{T_D_Hamiltonian})] and those obtained with the static Hamiltonian $H_{\mathrm{s}}$ [Eq.~(\ref{S_Hamiltonian})] for the LMG model. The maximum deviations obtained by comparing the quasienergies of $H(t)$ with the corresponding eigenvalues of $H_{\mathrm{s}}$ are represented by solid blue lines, while the maximum deviations obtained when comparing the mean values of $H(t)$ instead of the quasienergies are depicted by dashed orange lines. The difference between panels~a) and~b) is that in~a), the quasienergies have been restricted to the first Brillouin zone, whereas in~b), they have been extended to other zones when necessary. The vertical lines represent the values $T=1/\Omega=T_{\mathrm{c}}/4$ (dotted gray lines), which is the period value used in Section~\ref{Results}, $T=\pi/\Omega=\pi T_{\mathrm{c}}/4$ (dashed green lines), which is the value at which some eigenvalues of $H_{\mathrm{s}}$ start to exit the first Brillouin zone, and $T=4/\Omega$ (dash-dotted black line), which is the value of the characteristic time $T_{\mathrm{c}}$.} 
    \label{deviation}
\end{figure*}

It is important to note that the previous condition does not guarantee that all eigenvalues of $H_{\mathrm{s}}$ are necessarily within the first Brillouin zone. In fact, it is easy to see that it is always possible to ensure that none of these eigenvalues lies within that zone, without modifying the system's dynamics, simply by adding a constant term to the Hamiltonian $H(t)$. A sufficient condition for all eigenvalues of $H_{\mathrm{s}}$ to be within the first Brillouin zone can be obtained by imposing that $\lVert H_{\mathrm{s}}\rVert_2 \leq \pi\hbar /T$ for all values of $t _0$. Since $\lVert H_{\mathrm{s}} \rVert_2 \leq \max(\lVert H_1 \rVert_2, \lVert  H_2 \rVert_2)=\max(\lvert E_{1,\mathrm{M}}\rvert, \lvert E_{1,\mathrm{m}}\rvert,\lvert E_{2,\mathrm{M}}\rvert, \lvert E_{2,\mathrm{m}}\rvert)$, an alternative sufficient condition is given by $T/T_{\mathrm{c}}^{\prime}\leq \pi$,  with the characteristic time
\begin{equation}
	T_{\mathrm{c}}^{\prime} 
	=\frac{\hbar}{\max(\lvert E_{1,\mathrm{M}}\rvert, \lvert E_{1,\mathrm{m}}\rvert,\lvert E_{2,\mathrm{M}}\rvert, \lvert E_{2,\mathrm{m}}\rvert)}.
\end{equation}
If the period $T$ is sufficiently small compared to the characteristic time $T_{\mathrm{c}}$ in Eq.~(\ref{tc}) but some eigenvalue of $H_{\mathrm{s}}$ lies outside the first Brillouin zone, it would be necessary to extend the corresponding quasienergy of the time-periodic Hamiltonian in Eq.~(\ref{T_D_Hamiltonian}) to the appropriate Brillouin zone in order to achieve a good agreement.

For the LMG model considered in Sections~\ref{PhysRea} and \ref{Results}, $H_1=\Omega S_z/(2 S)$ and $H_2 = - \Omega S_x^2/(\hbar S^2)$, where $\Omega$ is a constant with dimensions of frequency. Therefore, assuming that $S$ is even, we have that $E_{1,\mathrm{M}}-E_{1,\mathrm{m}}=E_{2,\mathrm{M}}-E_{2,\mathrm{m}}=\hbar \Omega $, and consequently $T_{\mathrm{c}}=4/\Omega $. It is worth noting that the results presented in Section~\ref{Results} are surprisingly good, despite having chosen a period $T$ that is only a quarter of the time $T_{\mathrm{c}}$. Furthermore, since $T_{\mathrm{c}}' = 1/\Omega$, a sufficient condition for all eigenvalues of $H_{\mathrm{s}}$ to lie within the first Brillouin zone is $\Omega T \leq \pi$, which is satisfied for the period $T = 1/\Omega$ used in Section~\ref{Results}.

To illustrate the aforementioned, we propose a technique to quantify the deviation of our calculations in relation to the static model. Firstly, we compute the quasienergies and the mean energies of the Floquet modes for various values of the period $T$ as a function of the quench time $t_0$. Then, we compare the results obtained with those provided by the static Hamiltonian in Eq.~(\ref{S_Hamiltonian}) for the same values of $T$ and $t_0$. Specifically,  for each considered value of $T$, we calculate  the differences $e_n(t_0)-E_n(t_0)$ for all states $n$ as a function of the quench time $t_0$, where $e_n$ represents either the quasienergies or the mean energies corresponding to the time-periodic Hamiltonian in Eq.~(\ref{T_D_Hamiltonian}), and $E_n(t_0)$ represents the eigenvalues of the static Hamiltonian in Eq.~(\ref{S_Hamiltonian}). We select the maximum value
\begin{equation}
\label{deviation1}
D(T)=\max_{\{n,t_0\}}\big[\left| e_n(t_0)-E_n(t_0)\right|\big],
\end{equation}
which corresponds to the maximum deviation with respect to the static LMG model, and we plot this value against  the dimensionless  period $T/T_{\mathrm{c}}$ (Fig.~\ref{deviation}). The blue solid lines in both panels of Fig.~\ref{deviation} represent the maximum deviation of quasienergies, while the orange dashed lines correspond to the mean energies. It is worth noting that for $T\geq\pi/\Omega=\pi T_{\mathrm{c}}/4$, there are eigenvalues of the static Hamiltonian $H_{\mathrm{s}}$ that do not belong to the first Brillouin zone. For this reason, if the quasienergies are restricted to the first Brillouin zone, a discontinuity appears at $T/T_{\mathrm{c}}=\pi/4$ (dashed green vertical lines), as can be observed in panel~a) of Fig.~\ref{deviation}. This discontinuity arises due to the imposed restriction on the quasienergies. In fact, if we extend the corresponding quasienergies to the appropriate Brillouin zone, the discontinuity disappears, as can be observed in panel~b) of Fig.~\ref{deviation}.
The other two vertical lines correspond to the period $T=1/\Omega$ used in Section~\ref{Results} (dotted gray lines) and the characteristic time $T_c$ (black dot-dashed lines). As we can see in Fig.~\ref{deviation}, the quasienergies are closer to the calculations for the static model than the mean energies, but still the maximum deviation of both quantities for the selected period $T=1/\Omega$ are around $10^{-5}$--$10^{-4}$.

\section{\label{app:othermodels} Atom--diatomic molecule coexistence model}

\begin{figure}
\centering{\includegraphics[width=0.9\linewidth]{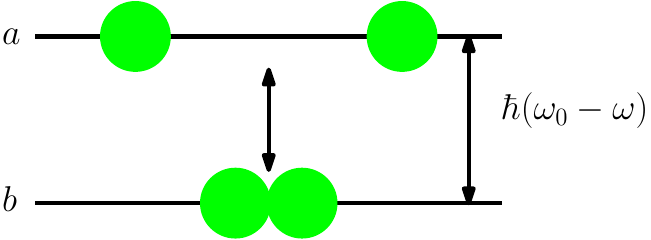}}
\caption{Schematic representation of the atom-diatom model. In this model, diatomic molecules ($b$) occupy the lower energy level, while single atoms ($a$) reside in the upper energy level. The energy difference, denoted as $\hbar(\omega_0 - \omega)$, corresponds to the energy required to dissociate the molecule into its two constituent atoms. This figure has been adapted from \cite{PF-2011}.
}
\label{app:fig-1}
\end{figure}

This work presents a new idea, which suggests that the same information regarding phase diagrams can be derived from either time-independent Hamiltonians or time-dependent quenched Hamiltonians. This concept is demonstrated in Sections~\ref{PhysRea} and \ref{Results} using the LMG model as an example. However, in order to emphasize that this concept is not limited to the LMG model, an additional example is provided here, which involves a simple two-level model describing the interaction between individual atoms and diatomic homo-nuclear molecules. Figure~\ref{app:fig-1} illustrates a schematic representation of this model.
\begin{figure}
    \centering
    \includegraphics[scale=0.15]{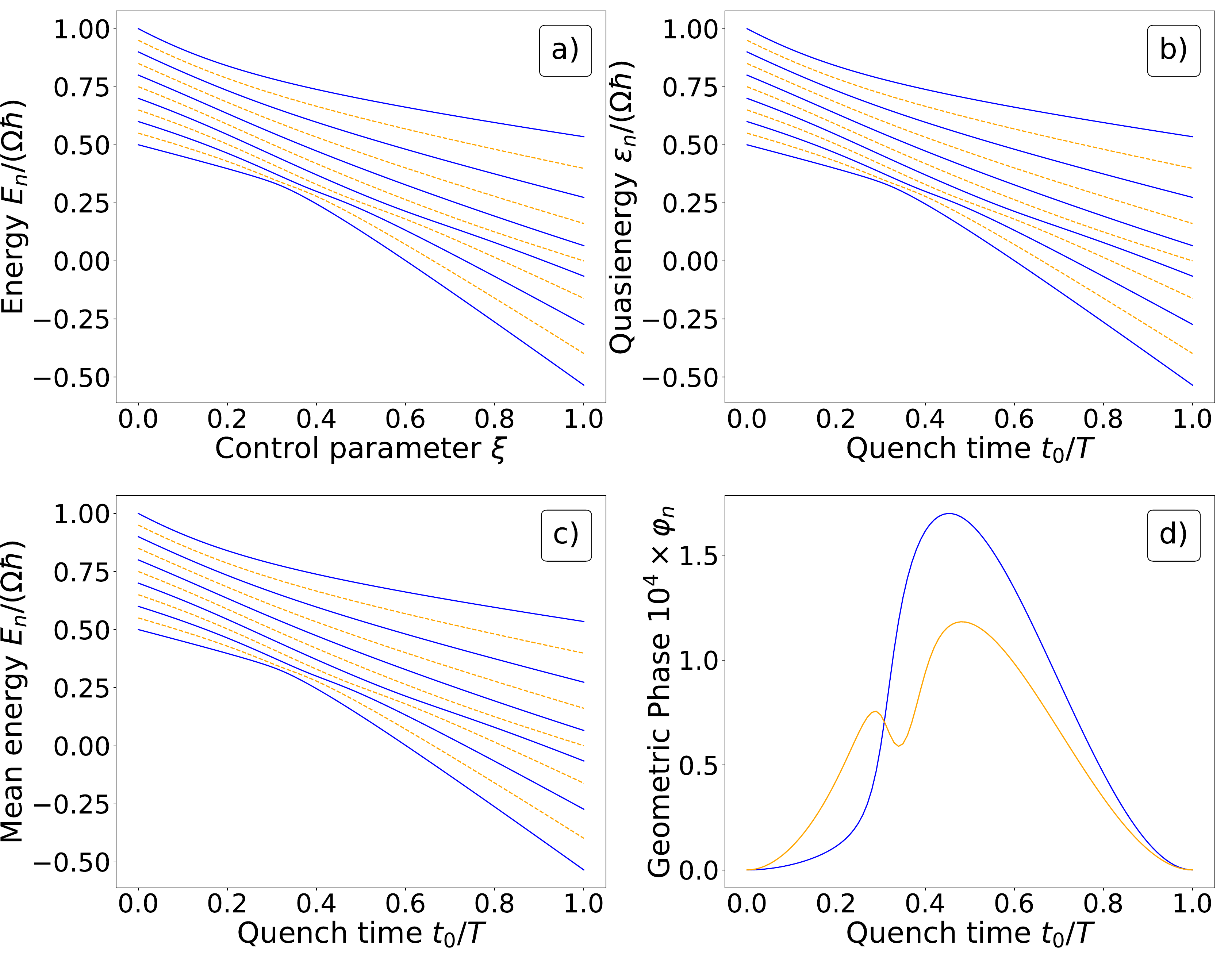}
    \caption{Energy correlation diagram for the atom-diatom model [Eqs.~(\ref{dia1}) and (\ref{dia2})]. Panel a) displays the correlation energy diagram for the time-independent Hamiltonian in Eq.~(\ref{Hamiltonian_CQPT}) as a function of  the control parameter $\xi$. In panel b), the correlation energy diagram is shown for the quasienergies in the Floquet time-dependent formalism. Panel c) illustrates the correlation energy diagram for the Floquet mean energies. Finally, the Floquet geometric phases for the ground and the first excited states are displayed in panel d). All calculations are performed with $M=20$, $\omega_0=2\Omega$, and $\omega=\Omega$.}
    \label{fig:atom-diatom}
\end{figure}

To describe the model using a bosonic formulation, two types of bosons are introduced: $a$, representing individual atoms, and $b$, representing diatomic molecules. The energy of each individual atom is $\hbar \omega_0/2$, while the energy of the diatomic molecule is $\hbar \omega$. These particles interact, and there are two limiting situations considered in Ref.~\cite{PF-2011} with Hamiltonians
\begin{equation}
  H_1 =  \frac{\hbar}{2M}\left(\omega_{0} n_a +2\omega n_b\right),
  \label{dia1}
\end{equation}
where $n_a=a^{\dagger}a$ is the particle number operator for type $a$-bosons (atoms) and $n_b=b^{\dagger}b$ is the particle number operator for type $b$-bosons (diatomic molecules), and
\begin{equation}
    H_2 = \frac{\hbar\Omega}{M^{3/2}}\left(b^{\dagger}aa+ba^{\dagger}a^{\dagger}\right) ,
    \label{dia2}
\end{equation}
where $M = n_a + 2n_b$ represents the total number of atoms, which is conserved,  and $\Omega$ is the coupling strength.

The time-independent energy correlation diagram of the model can be easily obtained using the Hamiltonian in Eq.~(\ref{Hamiltonian_CQPT}). In Fig.~\ref{fig:atom-diatom} panel a), the energy correlation diagram for the time independent Hamiltonian in Eq.~(\ref{Hamiltonian_CQPT}), as a function of the continuous control parameter $\xi$, is depicted. 
Following the idea presented in Section~\ref{Floquet-Theory}, the same problem can also be addressed using the time-dependent Hamiltonian of Eq.~(\ref{T_D_Hamiltonian}).
The results for the Floquet quasienergies and mean energies, as functions of the quench time $t_0$, are presented in Fig.~\ref{fig:atom-diatom}---panels b) and c), respectively--- for a period $T=1/\Omega$. In this case, $T_{\mathrm{c}}'=1/\Omega$ and the threshold value in Eq.~\eqref{tc} is $T_{\mathrm{c}}\approx 5.4695\Omega$, satisfying for our choice of $T$ the two conditions $T/T_{\mathrm{c}}^{\prime}\leq \pi$ and $T/T_{\mathrm{c}}$ sufficiently small. Despite the distinct frameworks employed, the similarities between the panel a) (time-independent) and panels b) and c) (time-dependent) are evident, as also observed in Fig.~\ref{fig-2}  for the LMG model. To provide a comprehensive overview, the Floquet geometric phases for the ground and the first excited  states are presented in panel d).


In order to verify the applicability of these ideas across different branches of physics, we conducted calculations for the following additional models, in addition to the one presented in Sections~\ref{PhysRea} and \ref{Results}, and earlier in this Appendix:
\begin{itemize}
\item An extended Lipkin model (see Ref.~\cite{Vidal-2006}), which exhibits a first-order quantum phase transition in its ground state. The phase diagram of this model resembles that of the Interacting Boson Model, which is of interest in nuclear physics.
\item The 2D limit of the vibron model (see Ref.~\cite{Curro-2008}), which is relevant in the study of the molecular bending spectra.
\item The Tavis-Cummings model (see Ref.~\cite{TC-1968}), which describes the interaction between radiation and matter and is of interest in quantum optics.
\end{itemize}
For all these cases, the results obtained provide further support for the conclusions reported in this paper.

\bibliographystyle{quantum}
\bibliography{refs}

\begin{thebibliography}{10}

\bibitem{PT_textbook}
H.E. Stanley.
\newblock ``Introduction to phase transitions and critical phenomena''.
\newblock \href{https://dx.doi.org/}{International series of monographs on
  physics}. Oxford University Press. ~(1987).

\bibitem{PT_textbook2}
H.~Nishimori and G.~Ortiz.
\newblock ``Elements of phase transitions and critical phenomena''.
\newblock
  \href{https://dx.doi.org/10.1093/acprof:oso/9780199577224.001.0001}{Oxford
  University Press}. ~(2010).

\bibitem{Sachdev2011}
S.~Sachdev.
\newblock ``Quantum phase transitions''.
\newblock \href{https://dx.doi.org/10.1017/cbo9780511973765}{Cambridge
  University Press}. ~(2011).

\bibitem{Vojta_2003}
M.~Vojta.
\newblock ``Quantum phase transitions''.
\newblock \href{https://dx.doi.org/10.1088/0034-4885/66/12/R01}{Rep. Prog.
  Phys. {\bf 66}, 2069}~(2003).

\bibitem{Carr2010}
L.~Carr, editor.
\newblock ``Understanding quantum phase transitions''.
\newblock \href{https://dx.doi.org/10.1201/b10273}{{CRC} Press}. ~(2010).

\bibitem{Cejnar2021}
P.~Cejnar, P.~Str{\'{a}}nsk{\'{y}}, M.~Macek, and M.~Kloc.
\newblock ``Excited-state quantum phase transitions''.
\newblock \href{https://dx.doi.org/10.1088/1751-8121/abdfe8}{J.\ Phys.\ A:
  Math. Theor. {\bf 54}, 133001}~(2021).

\bibitem{FloquetPedroPRL}
V.~M. Bastidas, P.~P\'erez-Fern\'andez, M.~Vogl, and T.~Brandes.
\newblock ``Quantum criticality and dynamical instability in the kicked-top
  model''.
\newblock \href{https://dx.doi.org/10.1103/PhysRevLett.112.140408}{Phys. Rev.
  Lett. {\bf 112}, 140408}~(2014).

\bibitem{FloquetPedroPRA}
V.~M. Bastidas, G.~Engelhardt, P.~P{\'{e}}rez-Fern{\'{a}}ndez, M.~Vogl, and
  T.~Brandes.
\newblock ``Critical quasienergy states in driven many-body systems''.
\newblock \href{https://dx.doi.org/10.1103/physreva.90.063628}{Phys. Rev. A
  {\bf 90}, 063628}~(2014).

\bibitem{rodriguez-vega2021}
M.~Rodriguez-Vega, M.~Vogl, and G.~A. Fiete.
\newblock ``Low-frequency and {M}oir{\'{e}}{\textendash}{F}loquet engineering:
  A review''.
\newblock \href{https://dx.doi.org/10.1016/j.aop.2021.168434}{Ann. Phys. {\bf
  435}, 168434}~(2021).

\bibitem{zhou_floquet_2021}
L.~Zhou and Q.~Du.
\newblock ``Floquet dynamical quantum phase transitions in periodically
  quenched systems''.
\newblock \href{https://dx.doi.org/10.1088/1361-648X/ac0b60}{J. Phys.: Condens.
  Matter {\bf 33}, 345403}~(2021).

\bibitem{deger_arresting_2022}
A.~Deger, S.~Roy, and A.~Lazarides.
\newblock ``Arresting classical many-body chaos by kinetic constraints''.
\newblock \href{https://dx.doi.org/10.1103/PhysRevLett.129.160601}{Phys. Rev.
  Lett. {\bf 129}, 160601}~(2022).

\bibitem{zhao_probing_2022}
S.~K. Zhao, Z.-Y. Ge, Z.~Xiang, G.~M. Xue, H.~S. Yan, Z.~T. Wang, Z.~Wang,
  H.~K. Xu, F.~F. Su, Z.~H. Yang, H.~Zhang, Y.-R. Zhang, X.-Y. Guo, K.~Xu,
  Y.~Tian, H.~F. Yu, D.~N. Zheng, H.~Fan, and S.~P. Zhao.
\newblock ``Probing operator spreading via {Floquet} engineering in a
  superconducting circuit''.
\newblock \href{https://dx.doi.org/10.1103/PhysRevLett.129.160602}{Phys. Rev.
  Lett. {\bf 129}, 160602}~(2022).

\bibitem{PRB2022Jangjan}
M.~Jangjan, L.~E.~F. Foa~Torres, and M.~V. Hosseini.
\newblock ``Floquet topological phase transitions in a periodically quenched
  dimer''.
\newblock \href{https://dx.doi.org/10.1103/physrevb.106.224306}{Phys. Rev. B
  {\bf 106}, 224306}~(2022).

\bibitem{LIPKIN1965188}
H.~J. Lipkin, N.~Meshkov, and A.~J. Glick.
\newblock ``Validity of many-body approximation methods for a solvable model:
  ({I}). {E}xact solutions and perturbation theory''.
\newblock
  \href{https://dx.doi.org/http://dx.doi.org/10.1016/0029-5582(65)90862-X}{Nucl.
  Phys. {\bf { 62}}, 188}~(1965).

\bibitem{Meshkov1965199}
N.~Meshkov, A.~J. Glick, and H.~J. Lipkin.
\newblock ``Validity of many-body approximation methods for a solvable model:
  ({II}). {L}inearization procedures''.
\newblock
  \href{https://dx.doi.org/http://dx.doi.org/10.1016/0029-5582(65)90863-1}{Nucl.
  Phys. {\bf { 62}}, 199}~(1965).

\bibitem{Glick1965211}
A.~J. Glick, H.~J. Lipkin, and N.~Meshkov.
\newblock ``Validity of many-body approximation methods for a solvable model:
  ({III}). {D}iagram summations''.
\newblock
  \href{https://dx.doi.org/http://dx.doi.org/10.1016/0029-5582(65)90864-3}{Nucl.
  Phys. {\bf { 62}}, 211}~(1965).

\bibitem{Grifoni1998a}
M.~Grifoni and P.~H{\"a}nggi.
\newblock ``Driven quantum tunneling''.
\newblock
  \href{https://dx.doi.org/https://doi.org/10.1016/S0370-1573(98)00022-2}{Phys.
  Rep. {\bf 304}, 229}~(1998).

\bibitem{Richerme2014}
P.~Richerme, Z.-X. Gong, A.~Lee, C.~Senko, J.~Smith, M.~Foss-Feig,
  S.~Michalakis, A.V. Gorshkov, and C.~Monroe.
\newblock ``Non-local propagation of correlations in quantum systems with
  long-range interactions''.
\newblock \href{https://dx.doi.org/10.1038/nature13450}{Nature {\bf 511},
  198}~(2014).

\bibitem{Jurcevic2014}
P.~Jurcevic, B.~P. Lanyon, P.~Hauke, C.~Hempel, P.~Zoller, R.~Blatt, and C.~F.
  Roos.
\newblock ``Quasiparticle engineering and entanglement propagation in a quantum
  many-body system''.
\newblock \href{https://dx.doi.org/10.1038/nature13461}{Nature {\bf 511},
  202}~(2014).

\bibitem{Frankbook}
A.~Frank and P.~Van Isacker.
\newblock ``Algebraic methods in molecular and nuclear structure physics''.
\newblock \href{https://dx.doi.org/}{John Wiley and Sons, New York}. ~(1994).

\bibitem{Dusuel2004}
S.~Dusuel and J.~Vidal.
\newblock ``Finite-{S}ize {S}caling {E}xponents of the
  {L}ipkin-{M}eshkov-{G}lick {M}odel''.
\newblock \href{https://dx.doi.org/10.1103/PhysRevLett.93.237204}{Phys. Rev.
  Lett. {\bf 93}, 237204}~(2004).

\bibitem{Romera2014}
E.~Romera, M.~Calixto, and O.~Casta{\~{n}}os.
\newblock ``Phase space analysis of first-, second- and third-order quantum
  phase transitions in the {L}ipkin{\textendash}{M}eshkov{\textendash}{G}lick
  model''.
\newblock \href{https://dx.doi.org/10.1088/0031-8949/89/9/095103}{Phys. Scr.
  {\bf 89}, 095103}~(2014).

\bibitem{Ribeiro2008}
P.~Ribeiro, J.~Vidal, and R.~Mosseri.
\newblock ``{Exact spectrum of the Lipkin-Meshkov-Glick model in the
  thermodynamic limit and finite-size corrections}''.
\newblock \href{https://dx.doi.org/10.1103/PhysRevE.78.021106}{Phys. Rev. E
  {\bf 78}, 021106}~(2008).

\bibitem{DAlessio2013}
Luca D’Alessio and Anatoli Polkovnikov.
\newblock ``Many-body energy localization transition in periodically driven
  systems''.
\newblock
  \href{https://dx.doi.org/https://doi.org/10.1016/j.aop.2013.02.011}{Ann.
  Phys. {\bf 333}, 19}~(2013).

\bibitem{hall-2015}
B.~J. Hall.
\newblock ``{Lie Groups, Lie Algebras, and Representations}''.
\newblock \href{https://dx.doi.org/10.1007/978-3-319-13467-3}{Springer New
  York}. ~(2015).

\bibitem{horn-1985}
R.~A. Horn and C.~R. Johnson.
\newblock ``Matrix analysis''.
\newblock \href{https://dx.doi.org/10.1017/cbo9780511810817}{Cambridge
  University Press}. ~(1985).

\bibitem{PF-2011}
P.~P\'erez-Fern\'andez, P.~Cejnar, J.~M. Arias, J.~Dukelsky, J.~E.
  Garc\'{\i}a-Ramos, and A.~Rela\~no.
\newblock ``Quantum quench influenced by an excited-state phase transition''.
\newblock \href{https://dx.doi.org/10.1103/PhysRevA.83.033802}{Phys. Rev. A
  {\bf 83}, 033802}~(2011).

\bibitem{Vidal-2006}
J.~Vidal, J.~M. Arias, J.~Dukelsky, and J.~E. Garc\'{\i}a-Ramos.
\newblock ``Scalar two-level boson model to study the interacting boson model
  phase diagram in the casten triangle''.
\newblock \href{https://dx.doi.org/10.1103/PhysRevC.73.054305}{Phys. Rev. C
  {\bf 73}, 054305}~(2006).

\bibitem{Curro-2008}
F.~P\'erez-Bernal and F.~Iachello.
\newblock ``Algebraic approach to two-dimensional systems: Shape phase
  transitions, monodromy, and thermodynamic quantities''.
\newblock \href{https://dx.doi.org/10.1103/PhysRevA.77.032115}{Phys. Rev. A
  {\bf 77}, 032115}~(2008).

\bibitem{TC-1968}
M.~Tavis and F.~W. Cummings.
\newblock ``Exact {S}olution for an {$N$}-{M}olecule---{R}adiation-{F}ield
  {H}amiltonian''.
\newblock \href{https://dx.doi.org/10.1103/PhysRev.170.379}{Phys. Rev. {\bf
  170}, 379--384}~(1968).

\end{thebibliography}

\end{document}